# Kondo effect by controlled cleavage of a single molecule contact.

R. Temirov[1], A. Lassise[1], F.B. Anders[2], F.S. Tautz[1]

[1] *Jacobs University Bremen, School of Engineering and Science, P.O. Box 750561, 28725 Bremen, Germany*

[2] *Universität Bremen, Institut für Theoretische Physik, P.O. Box 330440, 28334 Bremen, Germany*

Conductance measurements of a molecular wire, contacted between an epitaxial molecule-metal bond and the tip of a scanning tunneling microscope, are reported. Controlled retraction of the tip gradually de-hybridizes the molecule from the metal substrate. This tunes the wire into the Kondo regime in which the renormalized molecular transport orbital serves as spin impurity at half filling and the Kondo resonance opens up an additional transport channel. Numerical renormalization group simulations suggest this type of behavior to be generic for a common class of metal-molecule bonds. The results demonstrate a new approach to single-molecule experiments with atomic-scale contact control and prepare the way for the ab initio simulation of many-body transport through single-molecule junctions.

## 1. Introduction

One of the foremost challenges in the field of molecular electronics is the insufficient structural and electronic contact definition in many of the single-molecule transport experiments reported up to date. This demands the performance and/or statistical analysis of hundreds of experiments to extract the relevant data [1-3]. It has long been recognized that the scanning tunnelling microscope (STM) has the potential to make a difference in this respect [4-9]. However, two terminal STM transport experiments often lack the tunability which, for example, can be achieved when molecular junction is combined with a gate electrode [10-12]. Tunability of the transport junction and structural control are thus two important objectives which ideally should be realized in one and the same experiment, because this combination provides an optimal interface to *ab initio* transport calculations. Here we present a mechanically gated single-molecule transport experiment of quite general applicability in which tunability and contact control both reach a very high level.

Our approach to single-molecule transport experiments is based on perfectly ordered, epitaxial layers of molecules on a single-crystalline metal substrate. In such layers the electronic properties of the metal-molecule contact can be characterized by a wide range of spectroscopic experiments. Our experiments have been carried out on the PTCDA (4,9,10-perylene-tetracarboxylic-dianhydrid) molecule (Fig. 1A). The epitaxial PTCDA/Ag(111) interface [13, 14] is a model organic/metal interface, for which a large number of integrating [15-17] and single-molecule spectroscopies [18-20] as well as structural probes [21-24] have revealed a chemisorptive substrate-bonding of PTCDA through the π-electrons of its perylene core. Fig. 1A summarize the PTCDA-Ag contact: the former Lowest Unoccupied Molecular Orbital (shown as $L_0$ in Fig.1B) hybridises with metal states and is pulled below the Fermi level $E_F$ [15, 17, 18], yielding an occupation number close to two. In addition, there is a second interaction mechanism which involves the carboxylic oxygen atoms [22]. The two interactions correspond to distinct functionalities of the PTCDA molecule towards Ag(111) surface (delocalised π-electron system vs. carboxylic oxygen atoms). The separate bonding distance optimisation of these two parts above the Ag(111) substrate leads to a buckling distortion of PTCDA on Ag(111) [22] (schematically shown in Fig. 1A).

In our experiment, the molecular wire is established by a chemically bonded point contact between the epitaxially adsorbed PTCDA molecule and the tip. For this purpose, the molecule should possess a functional group which is free to react with the tip. Because of its two interaction channels towards metals, PTCDA is well-suited for this experiment, since it is in fact possible to transfer the bond of one of the four carboxylic oxygen atoms from the substrate to the tip and thus establish a molecular wire with well defined contacts (Fig. 1C).

After tip-molecule contact formation, the molecular wire junction can be stretched by retracting the tip, leading to a cleavage of the molecule-substrate bond and a simultaneous gating of the molecular wire. In the present example of the tip-PTCDA-Ag(111) wire, the transport spectra thus recorded not only reveal minute details about the metal-molecule bond breaking process, but also about the many-body transport physics in the molecular wire. To explain the observed behaviour, we suggest a scenario based on the Kondo effect [25, 26], which has been studied extensively in single-electron transistors [27-30], carbon nanotubes [31, 32], magnetic adatoms [33-37] and single molecules [10, 11, 38], to explain the results of our transport spectroscopy.

The paper is organized as follows: In section 2 we briefly summarize our experimental methods. In section 3 we will argue in detail that it is indeed possible to contact PTCDA as shown schematically in Fig. 1C. In section 4 we report the transport spectra of the mechanically gated molecular wire junction and discuss the implications of these data. In section 5 we discuss model calculations which allow one to rationalize the observed transport behaviour in terms of many-body correlations. Finally, the summary of the bond cleaving and transport scenario emerging from our experiment on structurally defined PTCDA wire junctions concludes the paper.

## 2. Experimental Methods

Experiments have been carried out in a low temperature STM (Createc GmbH) in ultra high vacuum at 10 K. The STM is located in a special ultra-low vibration laboratory. Electrochemically etched tungsten tips have been used for all experiments reported here. Before the measurements, the tips were annealed in-situ by electron bombardment. Between the measurements, tips were additionally prepared in the STM at low temperature by applying voltage pulses and by tip indentation into the clean Ag surface.

The Ag(111) surface was prepared by consecutive cycles of $Ar^+$ sputtering with an incident ion beam energy of 800 eV, followed by thermal annealing at 820 K. Prior to the deposition of PTCDA, the surface quality was controlled with Low Energy Electron Diffraction (LEED). PTCDA was deposited by sublimation from a home-built Knudsen cell, heated to 580 K. The substrate was held at room temperature during molecule deposition. The PTCDA material (commercial purity 99%) was purified by resublimation and outgassed in UHV for extended times (>100 hrs).

Two types of conductances are reported in this work. Firstly, we have measured approach and retraction spectra in which the current at low bias voltages (±2 mV) is recorded as a function of distance. These spectra yield linear conductances which are denoted as $I/V_b$. Because of the low bias voltage, this conductance very nearly corresponds to the zero bias value of the differential conductance $G=dI/dV_b$, i.e. $G(V_b{\approx}0){\approx}I/V_b$. Secondly, we have measured the differential conductance $G(V_b)$ at finite bias voltages, using a small bias modulation and lock-in detection. The modulation amplitude was 4 meV in most cases, while modulation frequencies are quoted in the respective figure captions.

The $z$-scale in our approach and retraction spectra is based on the calibrated piezo-voltage of our instrument. The calibration of the STM piezoelectric scanner was tested by measuring the height of mono-atomic Ag(111) steps. The result of 2.433 Å is in very good agreement with the expected value of 2.356 Å. In all approach and retraction curves the absolute value of the $z$-coordinate axis is referenced to the average position of the PTCDA carboxylic oxygen atoms. This scale has been calibrated in the following way. (1) The tip is stabilized at $I$ = 0.1 nA and $V_b$ = –340 mV ("stabilisation point") above the maximum of the PTCDA $L_0$ state. (2) With closed feedback loop, the tip is moved above bare silver, whence the tip moves (0.75 ± 0.02) Å toward to the surface. (3) The feedback loop is opened and the tip is moved into contact with the Ag surface. After an appropriate correction [6, 39], the corresponding $z$-piezo shift yields an absolute tip height above silver. From this absolute calibration and the knowledge of the adsorption height of carboxylic oxygen above PTCDA [22], a tip height of (6.7 ± 1.6) Å above the carboxylic oxygens of PTCDA at the stabilisation point is calculated. The uncertainty of 1.6 Å originates from considering two extreme (and unlikely) limits of tip-surface contacts. In reality, the (statistical) error bar will be much smaller; we estimate an error of ± 0.8 Å. A typical approach curve with the definition of the tip-metal contact point and a schematic of the complete $z$-calibration process are shown in the Supplementary Figure 1b and discussed there.

## 3. Mechanics of the Molecular Wire Junction

## 3.1. Contacting the Molecule

As mentioned in the introduction, the second contact necessary for our transport experiment as suggested in Fig. 1C is established as a chemical bond between the carboxylic function of surface-adsorbed PTCDA and the STM tip.

We have systematically investigated the possibility to form contacts to the molecule, by placing the tip at defined positions above the molecule and recording $I/V_b$-conductance profiles during tip approach and retraction. Several of such approach retraction curves are presented in Fig. 2A. When the approaching the tip above the π-system of PTCDA, an intermittent contact formation of the type previously reported for $C_{60}$ is observed [4, 8], while the C-H edge exhibits tunnelling-like behaviour up to the closest distances recorded here. In contrast, during approach at the carboxylic oxygen atom, the position of which is well known from systematic imaging experiments and STM simulations[18, 40], we observe abrupt current jumps which are beyond the resolution of the piezo-scanner. A typical approach-and-retraction curve measured over the carboxylic oxygens is shown in Fig. 2A (red curve). The two hysteretic current jumps in Fig. 2A can be explained by two successive jumps of the junction, first into and then out of contact, as schematically shown in Fig. 2B.

Abrupt contact jumps like the ones shown in Fig. 2A only occur in well-defined positions of the molecular layer as indicated by the constant current image displayed in Fig. 3A, which was performed at an average tip-to-sample distance approximately corresponding to the tunnelling conditions in the vicinity of the point 2 in Fig. 2A. The image reveals a set of sharp protrusions, each confined to a spatial region of ~1 Å diameter around the carboxylic oxygens of PTCDA. These features are due to the junction jumping first into and then out of contact while the tip travels laterally across the carboxylic oxygen atoms of PTCDA. The abrupt increase in current associated with the first jump leads the feed-back loop to retract the tip by ~2.5 Å, and the corresponding signal is thus shown as sharp protrusions in Figs. 3A,B. In Fig. 3A, only one type of carboxylic oxygen atoms react. However, if the tip is approached even closer, more carboxylic oxygen atoms respond in a similar way (Fig. 3B). From images such as Fig. 3A,B we must conclude that the contact between the tip and PTCDA is highly selective and, on the side of the molecule, involves a single atom.

We have also investigated the statistics of contact jumps, using a data set of 282 approach curves recorded at bias voltages of $\pm 2mV$. The approach-and-retraction curves themselves are shown in Fig. 5 and will be discussed in more detail below. Their statistics is displayed in Fig. 4. Just before switching (point 1 in Fig. 2A), the tunnelling conductance $I/V_b$ ranges between 0.002 and 0.02 $G_0$, where $G_0 = 2e^2/h$. Conductances of the junction after contact formation (point 2 in Fig. 2A) are an order of magnitude larger. Interestingly, it is even possible to resolve a difference in the response threshold of two types of molecules in the layer (Fig. 4B). Its well-known that in the commensurate superstructure of PTCDA on Ag(111) there are two non-equivalent sites [14, 18, 40]; the corresponding molecules are labelled A and B in Figs. 3A,B. As seen in Fig. 4B, type A molecules respond 0.2 Å later than type B molecules.

The next question which arises is what happens microscopically in the junction during the contact jump? From the properties of the molecule discussed in context with Fig. 1A, and the sharpness of the response as compared to approach at all other points, one may infer that the carboxylic oxygen atoms flip-up into contact with the tip, as shown schematically in Fig. 2B. Indeed, in the competition between the chemical $Ag_{substrate}$-O and the $Me_{tip}$-O interactions the latter must eventually become dominant, thereby attracting the oxygen to a new potential. The difference in the response thresholds mentioned above for the two molecules is in fact a strong indication that it is the carboxylic oxygen atoms which jump into contact. DFT calculations for the PTCDA/Ag(111) interface [22, 40] show that the height difference between the carboxylic oxygen atoms of the two molecules is of the order 0.05-0.1 Å and hence considerably smaller than 0.2 Å. It thus appears that the threshold distance in Fig. 4B does not scale with tip-oxygen distance, but rather with the oxygen-substrate bonding strength: in the case of type A molecules carboxylic oxygens are located on top of silver atoms, while in B molecule they are close to bridge sites. In our case the on-top geometry seems to result in a stronger bonding of the oxygens which in turn demands a closer approach of the tip. The different thresholds for A and B molecules thus corroborate the idea that the oxygen-substrate bond is in fact ruptured when the discontinuity in the current is observed: The oxygen atom flips up, and the current rises because it can then flow via the molecule, without the necessity to tunnel through space. The flipping up of the carboxylic oxygen atom was in fact also confirmed in a density functional calculation in which a tip was approached above the carboxylic oxygen atoms of PTCDA adsorbed on Ag(111) (cf. supplement).

We now discuss the nature of the tip-molecule contact. Once formed, it in principle allows the removal of individual PTCDA molecules from the surface by simple tip retraction (no current or voltage pulses are applied), as the artificial vacancy structure in the inset of Fig. 3C demonstrates. This contact strength rules out weak van der Waals forces between tip and molecule as basis of the contact. Moreover, the current discontinuity is independent of tip bias, which excludes purely electrostatic forces, too. The distribution of "after-contact" junction conductances, peaked around 0.03 $G_0$ and extending to 0.15 $G_0$ (see Fig. 4), also suggests a chemical bond [4, 8]. Finally tip-molecule contacts created by the STM tip and carboxylic function of PTCDA allow to pass currents of several microamperes similar to other chemical contacts [8]. In the Fig. 2B the blue curve represents one of the experiments where the tip molecule contact breaks at ~ -0.5 V, at this point the current through the junction has reached 5 μA. An image recorded immediately after the contact had broken (inset in Fig. 2B) reveals the destruction of the contacted molecule. All these observations, as well as the chemical specificity mentioned above, indicate that the tip-PTCDA contact is a chemical bond involving the carboxylic functionality of the molecule.

Concluding the discussion thus far, it is clear that an Ag(111)-adsorbed PTCDA molecule can be joined between metallic tip and substrate via two chemical contacts located at different functionalities of the molecule. The contact to the substrate ($\pi$-contact) is formed by the extended $\pi$-electrons of the perylene core and the carboxylic oxygen functionality, while the contact to the tip (T-contact) is formed via a single carboxylic oxygen atom as sketched in Fig. 1C. Because the two contacts are formed at different molecular functional groups which react as distinct entities with their respective bonding partners we reach the conclusion that the contacts are well-localised at their respective parts of the molecule, making this system a suitable model system for a single-molecule wire.

## 3.2. Stretching the Molecular Wire Junction

We now turn to a detailed discussion of the mechanical aspects of the junction stretching experiments. In total we have stretched several hundreds of junctions and simultaneously recorded the current at a fixed bias voltage of ±2mV. It has already been mentioned that the $Me_{tip}$-O bond is strong enough to remove molecules from the surface, by simple tip retraction. The success rate of such molecule manipulation, however, depends on the tip state, reaching more than 50% for some tips and falling below ~1% for others. Quite generally, all tips which we have used showed better success rates for PTCDA molecules residing at the edges of molecular islands or for completely isolated molecules on the Ag(111) surface. At the same time PTCDA molecules residing within compact islands were found to be the least prone to manipulation. This behaviour might in fact be explained by intermolecular interactions acting in the layer, which are known to have a strong influence on the structural and electronic properties of the layer [18, 40, 41].

At the end of an approach-retraction cycle, three possible outcomes of the contacted molecule may be distinguished (and are indeed observed in the various experiments): (1) After removal from the surface, the molecule may be transferred to the tip permanently, where in subsequent imaging experiments it yields contrast types which are well-known to result from a tip functionalised with PTCDA [40]; this was the case during the preparation of the artificial vacancy shown in the inset of Fig. 3C, which meant that after each removal the tip had to be cleaned above the bare Ag(111) surface. (2) After its *transient* removal from the surface, the molecule drops back to the surface and may still be imaged there, but at a different lateral position from where it was picked up. In these experiments the tip state stays preserved. Transient removal and preserved tips are often observed when molecules with reduced number of neighbours (e.g. isolated or island edge molecules) are contacted. Examples are shown below in section 4. (3) After the rupture of the $Me_{tip}$-O contact, the molecule can fall back into its old place on the surface. This often happens for early ruptures of the $Me_{tip}$-O contact and/or in compact layers where a lateral movement of the molecule is hindered by its neighbours. Similarly to the case (2), the tip state after such an event remains the same. Outcome (3) which quite reliably keeps the molecule and the tip states unaffected, allowed us to obtain a large statistical ensemble of junction experiments devoid of any dependence on the microscopic state of the tip (Figs. 4 and 5). In fact, Fig. 5 demonstrates a very good degree of reproducibility as long as the tip properties are unchanged.

To analyse more closely the mechanical changes in the junction occurring when the tip is retracted, we have plotted in Fig. 6 the *I/V* conductances of a few typical junction stretching experiments on a logarithmic scale, recorded on island (black and red), boundary (blue) as well as single (green) molecules, and ending with removal (red) or displacement (blue and green) of the molecule as well as its dropping back into the old position (black). In all cases, the jump-into-contact occurs at $z$-coordinates close to 3 Å. Within the first 2-3 Å of tip retraction, the conductances go through a maximum. The maximum is related to the transport physics of the molecular wire and will be analysed in detail in sections 4 and 5. Here, we concentrate on the general behaviour of the retraction spectra. For all curves, the current reaches the tunnelling level after retracting the tip by 10-12 Å. This is in remarkable agreement with the length of the molecule, which, measured between the nuclei of two carboxylic oxygen atoms, amounts to 11.5 Å. Evidently, this finding is a strong indication that the molecule remains in the junction until it is lifted up into an almost vertical position. Only at this point, one of the contacts, either to the substrate or the tip, finally breaks. We can thus be sure that in all cases in Fig. 6 we really lift up the molecule, even if after the experiment the molecule is found in its old position on the surface. Incidentally, the fact that the final rupture occurs after a tip retraction distance quite closely corresponding to the length of the molecule also indicates that the tip itself is mechanically stable during the junction stretching experiment.

Closer inspection of Fig. 6 reveals a further common feature of all curves. After approx. 4 Å retraction distance, the conductance curves, which in the range from 0 to 4 Å retraction distance show a smooth behaviour with conductance maximum, exhibit an – on overall decay which is superimposed with strong, abrupt jerks of the conductance. The conductance jerks indicate that at this stage the junction structure is not stable any more. Notably, the change always occurs after approximately the same stretching distance, which points to a distinct geometrical configuration of the junctions at which the instability occurs. The initial stages of the stretching have been analysed by a density functional calculation, in which after $Me_{tip}$-O contact formation the tip was retracted in steps of 0.2 Å. The simulation shows that during initial tip retraction the the part of the molecule closest to the tip is peeled off the surface by cleaving the π-contact to the surface. It is clear that due to geometric constraints, this peeling off must finally cross over into a regime in which the far end of the molecule slides over the surface while the molecule is lifted up further by the tip (cf. Fig. 9). The observed jerks and the corresponding junction instability in Fig. 6 may in fact be associated with this sliding motion.

## 4. Electronic Transport Spectra of the Molecular Wire Junctions

### 4.1 Experimental Results

We have seen in the previous sections that the mechanical strength of the tip-molecule contact allows us to tune and measure the electronic and transport properties of the molecular wire junction in wide limits, from equilibrium chemisorption of the molecule on Ag(111) to the point when electronic contact to the substrate is severed. It has been shown in the previous section that in 100% of the cases in which the junction was stable upon stretching (unlike the case shown in the Fig. 2A where the tip-molecule bond is broken at the early stages of tip retraction) the low-bias differential conductance $G=dI/dV_b$, evaluated as $G(V_b \approx 0) \approx I/V_b$, exhibits a surprising, yet well-pronounced maximum, as seen in Figs 5 and 6. Interestingly, the overall behaviour of the retraction spectra is similar in all three cases, apart from the systematic shift in conductance peak maxima between island, boundary or isolated molecules,.

A statistical evaluation of key parameters of the data set in Fig. 5 shows that the low bias peak conductance of the stretched junctions ranges between 0.1 and 0.3 $G_0$, which is a factor of approximately 8 larger than the conductance directly after jump into contact (Fig. 4A). In most cases, the peak conductance is reached after retracting the $z$-piezo by 1.5 Å, although there is also a notable broad distribution centred around 2.3 Å piezo travel (Fig. 4B). Following the analysis of the junction stretching mechanics in the previous section, it is clear that the conductance peak occurs while the molecule is peeled off the surface, before the junction instability sets in; it must therefore be associated with π-bond cleavage which is visualized by the DFT calculation shown in supplement.

To further investigate the origin of the conductance maxima in Figs. 5 and 6, we have performed a full spectroscopic characterization of 29 molecular junctions during stretching. This was achieved by interrupting the tip retraction at regular intervals to record full differential $G(V_b)$ conductance spectra using the lock-in technique. These experiments were carried out on molecules at island boundaries and separate molecules, because contacts to molecules with reduced number of neighbours were found to be more stable at the high current loads which are inevitable during the spectroscopic measurements. Two experiments with full spectroscopic data are displayed in Figs. 7A,C (junction I, recorded for a separate molecule) and B,D (junction II, recorded on a molecule on an island border). After both experiments, the surface region was imaged again (insets in Figs. 7A,B), indeed revealing the displacement of each contacted molecule from its original position. We note that the retraction spectra of junctions I and II have already been shown on a logarithmic scale in Fig. 6 (green and blue curves respectively).

The behaviour of differential conductance $G(V_b)$ upon junction stretching is summarized as follows: Immediately after contact formation the overall conductance jumps by an order of magnitude with respect to the tunnelling regime. In the early stages of tip retraction, changes of $G(V_b)$ are relatively small (blue series in Fig. 7C). Later the conductance at negative bias starts to increase, because a spectral feature approaches the Fermi level ($E_F$) from the left (green series in Fig. 7C). This process is accompanied by the simultaneous sharpening of the approaching peak. Remarkably, upon further stretching the peak stays pinned at $E_F$ and its FWHM does not change any more, while the peak intensity gradually decays (red series in Fig. 7C), until the junction conductance becomes unstable (Note that the pinned peak can be observed well into the regime where conductance jumps abruptly). The experiment on junction II exhibits a similar behaviour, the only difference being that the intermediate (green) series of peak sharpening and approach to $E_F$ is bypassed and an immediate transition from the low- to the high-conductance state, the latter characterized by the narrow peak at $E_F$, is observed instead. In fact, the gradual shift of the peak to the Fermi level represented by the green series in Fig. 7C is a generic property of stretching experiments on isolated molecules. The systematic differences between in stretching behaviour of isolated and island edge molecules depicted in the Figs. 7C,D might be due to the intermolecular interactions in the layer. In the following, we will refer to the blue and green spectra in Fig. 7C as belonging to *phase 1* of the stretching experiment, while the red spectra belong to *phase 2*. The regime in which the junction is unstable is referred to as *phase 3*.

## 4.2 Discussion

Firstly, we would like to point out that the experimental approach reported here offers a very good opportunity to measure transport on molecular wires *with well-defined electrode contacts*. In the field of molecular electronics, contact definition is a particular challenge, because on the one hand traditional single molecule transport experiments, e.g. mechanically controlled break junctions [42-44], present only very limited chances to structurally characterise, let alone control, lead contacts, while on the other hand theoretical calculations reveal that the contact properties have a strong influence on the current transport through the wire, sometimes even stronger than the molecule itself [1, 2].

In our experimental approach, one of the two contacts is formed to a single crystalline surface and hence may be characterised by the full range of surface science spectroscopies and microscopies. The other contact is to the tip of an STM. Of course, this contact is structurally less well-defined than the former, but it is still possible to determine exactly which part of the molecule reacts with the tip. This is particularly important for more complex molecules which contain more than one reactive group; without the controlled approach of the STM tip it would be impossible to select which of the functional groups actually forms the lead contact. Moreover, with systematic approach-and-retraction spectra, in the future possibly coupled with simultaneous force measurement, one can characterise this tip-molecule contact very well, before full current-voltage transport spectra are recorded.

We have seen above that in spite of our well controlled junction geometry, one still observes variations in behaviour which are currently beyond control, yet often show a systematic pattern (e.g. generic differences in stretching of island and isolated molecules). However, it is also true that the number of experiments which are needed to clarify the physics is much smaller than in traditional techniques, because one can from the outset eliminate a large number of "failures" by the careful preparation of the single crystalline sample surface, the molecular layer, and the tip. Moreover, with the imaging capability of the STM the *a priori* selection of appropriate objects becomes possible. Further developments of the technique may reduce the need for statistics even further.

We foresee that the experiments like the one reported here may be conducted on a larger number of different molecules. The only requirements of the technique are the existence of a well-defined adsorption geometry of the molecule on the surface and the existence of a dedicated functional group to which the tip can bind strongly enough to pick up the molecule. In all of such cases there are then good chances that a molecular wire may be established. Because the tip can be approached to the functional group which one wishes to contact, we are not limited to those functional groups like thiols which spontaneously form bonds to a contact in a self-organisation process.

With the method proposed here, the large body of knowledge accumulated over the last decades in the field of surface science can be brought to bear on molecular electronics in general and single molecule transport experiments in particular. Moreover, the approach will allow the *ab initio* theoretical simulation of transport in realistic systems the structure of which has been established by experiment. In a first step, *ab initio* geometric and electronic structure calculations are to be performed; if they recover the experimental geometry, the electronic structure thus obtained can then be used as the basis of transport calculation at various levels of sophistication, including the first-principles level.

Let us now turn to the particular results of the present experiments and try to understand the behaviour of the PTCDA wire under stretching. On the basis of our knowledge about the PTCDA/Ag(111) electronic structure, we can explain the behaviour in *phase 1* in Fig. 7 as being caused by the shift of the bonding orbital towards $E_F$, due to $\pi$-bond cleavage. In fact, the behaviour precisely follows standard chemisorption theory [45] *in the reverse*: As the molecule *approaches* the surface during the chemisorption process, the interaction between the chemisorbate-to-be and the surface causes a downshift of the molecular LUMO (affinity level), because it is stabilized by image forces. Eventually, the affinity level may end up below the Fermi level $E_F$ and be filled by charge transfer from the metal, as in the present case of PTCDA/Ag(111). Additionally, the molecular level will be broadened due to hybridisation with metal states. We note here that for PTCDA the molecular affinity level extends over the perylene core of the molecule. Filling and broadening of this level thus constitutes the $\pi$-bond (or $\pi$-contact) between in PTCDA and Ag(111) (Fig. 1A). From now on we call the bonding orbital corresponding to the $\pi$-contact $L_0$ and its energy $\varepsilon_0$. On stretching the junction and *cleaving* the $\pi$-contact, the chemisorption process is reversed: $L_0$ moves up in energy and sharpens again. This is precisely what is observed experimentally in Fig. 7, where the differential conductance $G(V_b)$ is determined by the changing density of states of the metal-molecule π-contact.

We note in passing that the mechanical strain on the junction here acts in a similar way as gate voltages in a single-electron transistor do, shifting the energy $\varepsilon_0$ of the electronic level $L_0$ with respect to $E_F$ of the contacts. Note, however, that the mechanical gating employed here simultaneously changes the coupling constant $\Gamma_S$ of the molecule to the surface.

In *phase 2*, we observe a deviation from the simple picture of reverse chemisorption. On its basis, we would naively expect a continued shifting of $L_0$ through the Fermi level to energies $\varepsilon_0 > 0$. This is, however, not what is observed. Rather, the conductance peak becomes pinned at $E_F$, as mentioned already above. At the same time, the standard theory of the interaction of local with extended states, as introduced by Anderson for impurities in bulk metals [46] or by Newns for the chemisorption problem [47], predicts that the simple single-particle picture of the bond cleaving process outlined in the previous paragraph only holds until the moment when the $L_0$ comes within a characteristic energy scale of the Fermi level, at which many-body correlations in the molecules and its contact come into play [48]. This energy scale is defined by electron correlations in the local orbital.

In the remainder of the paper we will explore the notion that the experimentally observed pinning and the electron correlation are related in the present case. It may be true that other mechanisms for the pinning are conceivable, or even that the pinning is purely accidental. But the point which we will argue below is that, on the appropriate energy scale, the effects of correlations are very likely to show up and cause pinning of the resonance $L_0$.

The Coulomb repulsion between electrons determines crucially the physical properties of atoms and molecules. In single-electron transport experiments, the fluctuation of the electronic occupation in a molecule is associated with a charging or correlation energy U due to the electron-electron interaction. For an extended orbital like PTCDA's $L_0$ in the presence of a well-screening metal surface, we may expect this energy to be small, but certainly not zero. If the level position of the orbital $L_0$ is far away from $E_F$ on the scale of $U$ (i.e. $|\varepsilon_0| >> U$), $L_0$ can be regarded as a doubly occupied singlet. If, however, the orbital is close enough to $E_F$ e.g. as the result of the mechanical gating in our experiment), the Coulomb repulsion U forbids such a doubly occupancy, and the singly occupied orbital carries a magnetic moment, if at the same time the contact coupling parameter $\Gamma$ is sufficiently weak ($U/\Gamma > 1$) to prevent the electron to become delocalized. Since the mechanical gating tunes both $\varepsilon_0$ and $\Gamma$ towards zero, with $U$ varying only weakly, we may indeed expect our wire to enter a regime in which many-particle correlations become relevant. In particular, Kondo-like correlations may arise between the singly occupied molecular level $L_0$ and the electrons in the two leads, i.e. substrate and tip, leading to a transport resonance pinned at the Fermi level [44].

## 5. Model Calculation

### 5.1 Introduction of the model

To confirm that the scenario sketched out above predicts a resonance in the differential conductance pinned at the Fermi level, we have carried out a simple model calculation in which the essential ingredients of our system are present: a local level $L_0$ which is shifted by some gating effect with respect to one electrode, effective couplings $\Gamma_S$ and $\Gamma_T$ between this level and the electrodes, and a correlation energy $U$ in the local level. This is the Anderson model, which – as pointed out by Newns – also captures the essential physics of the chemisorption problem. Of course, in principle one should take into account that transport through a molecular junction depends on contributions from all its different molecular orbitals, the details of their couplings to the substrate and the STM tip, and also the direct STM/substrate coupling. However, from the absence of Fano line-shapes in our *dI/dV* spectra we can exclude a direct tip-substrate tunnelling current [33-35, 49]. Moreover, the lack of a peak sub-structure in our data suggests that the major contribution to the transport current stems from charge transport through a single, strongly substrate- and tip-coupled orbital. As it stands, this can only be the level $L_0$ (see above). For the correlation (or local charging) energy $U$ in $L_0$ (proportional to $e^2/r$ ) a reasonable estimate of 100 meV follows from the spatial extension of the LUMO (~10 times a typical atomic radius) and the dielectric screening in the close vicinity of the metal.

We now discuss the basic features of our model calculation. The lead coupling of a single level is determined by two lead coupling functions $\Gamma_\alpha(\omega) = \pi \sum_k |V_{\alpha k}|^2 \delta(\omega - \varepsilon_{k\alpha})$, one each for the tip ($\alpha$=T) and the substrate ($\alpha$=S); details of individual coupling matrix elements $|V_{\alpha k}|^2$ to a conduction electron with momentum $k$ thus only enter in an averaged manner [50]. To leading order, these coupling functions can be replaced by energy-independent coupling constants $\Gamma_S$ and $\Gamma_T$. In this simplified model, the electrical current as function of bias is determined by the Laudauer-type formula [51]

$$I(V_b) = \frac{2e}{h} \frac{4\Gamma_S \Gamma_T}{(\Gamma_S + \Gamma_T)^2} \int d\omega \big[ f(\omega - \mu_T) - f(\omega + \mu_S) \big] T(\omega) \,, \qquad (1)$$

where the bias e$V_b$=$\mu_T$-$\mu_S$ is given by the difference between the chemical potentials of substrate and tip[52]. The zero bias differential conductance [51]

$$G(0) = \widetilde{G}_0 h(T/T_K) \qquad (2)$$

is obtained by differentiating equation (1) with respect to the bias. The universal function $h$(x) is a convolution of the *T*-matrix with the derivative of the Fermi function. Its argument is the reduced temperature $T/T_K$, where the Kondo temperature $T_K$ is a characteristic temperature below which many-body effects become relevant. In our numerical simulations we use a single parameter $\Gamma$ which is given by $\Gamma = \Gamma_S + \Gamma_T$.

## 5.2 Results of the model calculation

We have used the numerical renormalization group method (NRG) [50] to calculate the equilibrium $T$-matrix [53, 54] of our model Hamiltonian as function of energy $E$ and level position $\varepsilon_0$. In Figs. 8A,B the magnitude of the $T$-matrix is plotted as colour map. Two different parameters sets are displayed: Fig. 8A has $\Gamma_S + \Gamma_T = 10$meV and $T=1$K, while Fig. 8B has $\Gamma_S + \Gamma_T = 20$meV and $T=9$K. The $T$-matrix is proportional to the experimentally measured differential conductance. Accordingly, vertical cuts through the $(V_b, \varepsilon_0)$-plane correspond to differential conductance spectra such as plotted in Figs. 7C,D, while horizontal cuts would reveal how the differential conductance at fixed bias depends on the $L_0$ binding energy $\varepsilon_0$.

The plots in Fig. 8 show that for a doubly occupied ($n=2$) or an empty ($n=0$) level many-body effects indeed remain weak: the $T$-matrix is in these cases determined by a Lorentzian-broadened single-particle level located at either $\varepsilon_0 + U$ (for $n=2$) or $\varepsilon_0$ (for $n=0$). The first corresponds to the removal of one electron from the doubly occupied level $L_0$, the second to the addition of one electron to the empty $L_0$. Evidently, both processes allow charge transport through the molecule.

However, Fig. 8A also shows that as soon as soon as $L_0$ is shifted up so far that $\varepsilon_0 + U$ becomes equal to $E_F$, the orbital occupancy in $L_0$ is reduced to $n=1$ and the single level splits into two differential conductance peaks at $\varepsilon_0$ and $\varepsilon_0 + U$, seen as light blue features in the colour map [48]. Additionally, Figs. 8A,B clearly show a new resonance pinned close to $E_F$; this is a many-body resonance, and its origin is the finite correlation energy $U$ and the magnetic moment in the molecular orbital $L_0$ at odd-integer filling $(n=1)$ [48].

In Fig. 8A the many-body resonance is indeed accompanied by two well-resolved single-particle resonances at $\varepsilon_0$ and $\varepsilon_0+U$ (cf. the cut through Fig. 8A shown in Fig. 8C). This is the situation of the classical Kondo effect, where charge fluctuations in the local orbital $L_0$ are completely suppressed and the only remaining excitations associated with $L_0$ are spin-flips. But crucially, in accordance with Friedel's sum rule [55], our NRG simulation predicts the Fermi level pinning of the differential conductance peak even in the weakly correlated mixed valence regime of Fig. 8B, where $\Gamma$ is so large that charge fluctuations in the orbital $L_0$ are not completely suppressed. The resonance at $E_F$ in Fig.8B is then a superposition of the two single-particle peaks in the left and right tails and the many particle peak in the centre (cf. blue curve in fig. 8C). The model calculation thus reproduces the experimental behaviour qualitatively. In the next section we will analyse its relation to the experimental data in more detail.

## 5.3 Relation to experiment

We now relate the results of the model calculation to our experimental findings, and discuss separately the two phases of the experiment.

### 5.3.1. Phase 1

We first note that, in agreement with the NRG prediction for the even-integer occupancy regime, we do observe in Fig. 1B and the green series of Fig. 7C a single resonance in the differential conductance. This resonance persists as long as the overwhelming majority of DOS relating to $L_0$ is below the Fermi level, i.e. for $n$ close to 2.

Let us now turn to the pinning regime. As pointed out in the previous section already, both experiment and model calculation show this pinning qualitatively. But beyond this qualitative accordance, the simulation of fig. 8B and the experiments show a quantitative agreement if physically reasonable model parameters are assumed. Following Kondo theory, the width of the peak at $E_F$ is determined by the parameter ratio $U/\Gamma$, changing from $\Gamma$ at $U/\Gamma \ll 1$ to an exponentially small value

$$\Gamma_{eff} \propto \exp\left[\pi\varepsilon_0(\varepsilon_0 + U)/(2U\Gamma)\right] \qquad (3)$$

at $U/\Gamma \gg 1$ (in the limit $T \rightarrow 0$). In the model calculation of Fig. 8B, carried out at the experimental temperature 9 K and with $U/\Gamma = 5$, the width of the transport resonance decreases from 40 meV (FWHM) in the single-particle regime to 20 meV for the pinned peak. Experimentally, we observe a width of the pinned peak of 23 meV (junction I) and 13 meV (junction II). In the case of junction I, where we have access to the relevant regime, the width of the pinned peak corresponds to about *half* the width of the single-particle $L_0$ just before pinning occurs, in good agreement with the simulation of Fig. 8B.

Next to the absence of well-resolved, separate single-particle peaks at $\varepsilon_0$ and $\varepsilon_0 + U$, this experimentally observed moderate sharpening is further evidence that at the end of *phase 1* our wire is in a *weakly* correlated regime. We thus can conclude that the mechanical gating tunes our wires into borderline Kondo physics. From the peak width of the pinned resonances we can estimate Kondo temperatures in the range 80 K (junction II) to 160 K (junction I).

In principle, it would be desirable to test the proposed mixed valence/Kondo scenario in our junctions further by measuring transport spectra at temperatures above $T_K$, where the many-body resonance, and thus the pinning, should vanish [26]. However, it has turned out that such experiments are impossible. To record a series of spectra as in Fig. 7, the molecule needs to be stabilized in the junction for ~5 minutes, with the feed-back loop of the STM turned off. The experiment thus relies on extremely low drift levels (both laterally and vertically), which in our instrument cannot be achieved at temperatures above base temperature of 6-10 K, let alone at temperatures in excess of 150 K. Similarly, a common test for Kondo physics is the application of magnetic fields to split the Kondo peak [26]. But such high Kondo temperatures would need magnetic fields stronger then 100 T [26], which are clearly not available.

### 5.3.1. Phase 2

We now turn to the second phase of the stretching experiment, when the pinned peak decays on further stretching. The peak position remains close to chemical potential, and no re-appearance of a single, shifting peak for $V_b>0$ is observed. Instead, the experiments cross over into a regime in which the zero-bias differential conductance peak decays. This implies that the level position does not cross the chemical potential in *phase 2*.

On the one hand, it seems unlikely that our experiment can tune the level $L_0$ into the *n=0* regime (i.e. $\varepsilon_0>0$), because this would correspond to a free PTCDA molecule. But in our experiment the molecule must stay in contact with at least one of the electron reservoirs. This may well mean that molecular level does not become unoccupied (i.e. at least $\varepsilon_0$ remains below $E_F$). On the other hand, the major effect on the differential conductance in this regime is given increasingly asymmetric coupling of the molecular orbital to the substrate and the STM tip. Stretching the contact yields a reduction of the molecular coupling to the surface $\Gamma_S$ and therefore the total coupling $\Gamma = \Gamma_S + \Gamma_T$. The many-body resonance width is reduced, but remains pinned to the chemical potential close to the unitary limit. However, the conductance prefactor

$$\widetilde{G}_0 = \frac{2e^2}{h} \frac{4\Gamma_S\Gamma_T}{(\Gamma_S + \Gamma_T)^2}, \qquad\qquad (4)$$

reaches its maximum of *2e²/h* only for a symmetric junction $\Gamma_S = \Gamma_T$ at zero temperature and zero bias. At finite temperatures, the conductance will be reduced on a universal scale $T/T_K$, but theory still predicts a narrow peak of half-width $k_B T_K$ in the differential conductance at zero bias. Crucially, equation (4) predicts a decay of the pinned transport resonance, if the asymmetry between the two contact couplings $\Gamma_S$ and $\Gamma_T$ is increased. Evidently, this may be achieved in the process of removing the molecule from the substrate.

The data in Fig. 7 allow us to derive a few general tendencies concerning the behaviour of the two lead couplings $\Gamma_S$ and $\Gamma_T$ and thus the asymmetry term in equation (4) in phases 1 and 2 of the experiment. *(i) Directly after contacting the molecule with the tip, $\Gamma_T < \sim \Gamma_S$, i.e. the tip coupling is marginally smaller than the substrate coupling.* This can be inferred from the data in Fig. 1B, where the differential conductance of an unstrained junction immediately after contact formation (blue curve) is compared to the tunnelling spectrum recorded with the tip far from the molecule (red curve). By contacting, the FWHM of $L_0$ is increased from 300 to 400 meV. Using the relationship $\Delta = \sqrt{\Delta_S^2 + \Delta_T^2}$ and identifying $\Delta_S \approx 300$ meV with $\Gamma_S$ and $\Delta_T$ with $\Gamma_T$, we find $\Gamma_T < \sim \Gamma_S$. *(ii) At the beginning of phase 2 the junctions are already operated in a strongly asymmetric regime.* On the one hand, this follows from the peak heights in *phase 2* of the junction stretching experiment. While equation (4) predicts unitary conductance at zero bias for a symmetric junction, our experimental junctions show maximum differential conductances in the range 0.12 $G_0$ (junction I) and 0.08 $G_0$ (junction II). These are typical values also for other experimental junctions which are not shown. Since experiments are performed well below the Kondo temperature $T_K$, this observation proves that at beginning of phase 2 the junction is already asymmetric. The analysis of the peak heights with equation (4) reveals that in the case of junction II, the initial asymmetry in phase 2 is 80, rising to $\infty$ at the end of the experiment. On the other hand, the behaviour of the peak widths in *phase 2* also proves the above statement, if it is remembered that via equation (3) the width of the Kondo peak is determined by parameter $\Gamma = \Gamma_S + \Gamma_T$. Since the widths of the pinned peaks in the Figs. 7C,D do not change significantly as their intensity decays, we can again conclude that at the beginning of *phase 2* the junctions are already operated in a strongly asymmetric regime; otherwise one would expect the increasing asymmetry to be associated with a reduction of the overall $\Gamma$ and thus a significant peak sharpening. Maximum peak heights and peak width evolution thus present a consistent picture. *(iii) In phase 2, we expect $\Gamma_S << \Gamma_T$ to hold.* As the molecule is peeled off the surface in *phase 1*, $\Gamma_S$ must decrease, and given statement (i) above, the junction should therefore pass through a symmetric configuration relatively early, before it reaches *phase 2*. In *phase 2*, the asymmetry required by statement (ii) is then achieved by $\Gamma_S << \Gamma_T$.

We also note here that since the junctions are operated in a strongly asymmetric regime, non-equilibrium renormalizations of the *T*-matrix can be neglected in leading order *(5)*, and the NRG calculations remain valid even at small finite bias.

## 6. Summary and Conclusion

Finally, taking into account the discussion of the previous sections, the scenario sketched in Fig. 9 can be proposed for our junction stretching experiments: Initially, on tip approach the carboxylic oxygen atoms flip up and form a covalent bond with the tip which is mechanically strong enough to lift the molecule from the surface. Electronically, the tip-molecule contact is marginally weaker than the π-contact of the molecule to the substrate. We thus establish a single-molecule wire with two structurally well-defined lead contacts. When the STM tip is retracted, the wire junction is stretched and its transport properties evolve in three phases. In the first phase, the molecule-metal π-bond is cleaved. This process can be described in a single-particle picture as reversed chemisorption: The doubly occupied bonding orbital $L_0$ is shifted up in energy and sharpens. During this process, the molecule-substrate coupling $\Gamma_S$ is reduced to values below $\Gamma_T$. Eventually, the level comes so close to the Fermi level that its occupancy is reduced to $n \approx 1$ the molecule evolves from a non-magnetic chemisorbate to magnetic centre and a Kondo-like many-body resonance develops at the Fermi level. This marks the transition between phases 1 and 2 in Fig. 9. In our experiments, the many-body physics of the molecular wire is characterised by (1) weak correlations in the mixed valence regime and (2) a competition between the zero-bias $T$-matrix increasing towards unity (due to electron correlations) and a decrease of the pre-factor $\widetilde{G}_0$ (due to increasing asymmetry $\Gamma_S \ll \Gamma_T$). Therefore, the wire does not show the maximum conductance of the unitary limit. In phase 2, the wire remains in the $n \approx 1$ regime, but its contact asymmetry is further increased, up to the point when – in most of the investigated cases, including junction I and II – the electronically stronger contact to the tip is ruptured. After this rupture, the molecule falls back to the surface in a different position/orientation from before, showing that the extended bond to the molecule was indeed transiently cleaved.

All experimental data can be consistently explained in the above scenario. The physical parameters which can be estimated from the experiment, i.e. $\Gamma$ and $\varepsilon_0$, are consistent with the results of our model calculation, in the sense that for these parameters the model predicts key observables like line shape, height, and width of the many-body resonance at the Fermi level in near-quantitative agreement with experiment, if a physically reasonable estimate for $U$ is employed. But let us stress again that we cannot exclude the possibility that the pinned resonance at the Fermi level has another origin. However, the point is that if another mechanism is present, the correlation scenario discussed here can be expected to contribute to the behaviour of the wire, even if it is dominated by another, as yet unknown mechanism. If on the other hand, our scenario is indeed the right one, presented experiment is one of the first observations of Kondo physics in a molecule without magnetic ion [44, 56].

The behaviour observed here should not be specific to PTCDA; rather, we expect it to be a general occurrence whenever a chemisorption bond of the type considered here is cleaved, or indeed formed, and the bonding orbital is shifted through the Fermi level, a standard situation in the theory of chemisorption. In such cases, the present experiment in principle offers a sensitive way to experimentally determine correlation energies in the relevant molecular orbitals. Furthermore, our experiment clearly shows the importance of electron correlations in the microscopic dynamics adsorption process itself, which may be relevant for the theoretical determination of adsorption potential energy surfaces and sticking coefficients.

We anticipate that our transport experiments, and others of similar kind on different molecules and substrates, will in future allow the detailed comparison with non-equilibrium many-body transport and calculations at the *ab initio* level, on the basis of the precise geometrical and electronic junction structure. On a more general note, the approach taken here may be a new pathway to well-defined single-molecule experiments. In particular, our experiments combine efforts to symmetrize the STM junction for transport experiments with optimal contact control by using molecules from epitaxial layers.

**Figure 1 | Structural and electronic properties of PTCDA/Ag(111). A,** Structural formula of PTCDA and schematic side view of PTCDA bonding to Ag(111), with primary $\pi$–interactio and secondary interactions via the carboxylic oxygen atoms (vertical distortion is shown schematically). Yellow arrows: Charge transfer from metal to LUMO due to primary bond. **B,** Tunnelling spectrum (red, tip-to-PTCDA distance > 4 Å) and transport spectrum after tip-O contact formation (blue, tip-to-PTCDA distance ≈ 2 Å). The discontinuity in latter is caused by destruction of the molecule (cf. image in the inset) by $I \approx 5$ µA. White cross in the inset indicates the position of the approach. Both spectra were measured with lock-in detection using the modulation of 4 mV and 500-1000 Hz **C,** Schematic side view of the junction with the PTCDA molecule incorporated between the STM tip and the surface.

**Figure 2 | Forming a covalent contact between the STM tip and PTCDA. A,** Approach spectra recorded above various parts of PTCDA (left ordinate axis) and Ag(111) (right ordinate axis), as labelled, recorded with $V_b$ = 2 mV (dark: approach, light: retraction, besides the red curve where time evolution goes along increasing numbers). $I/V_b$-conductance in units of $G_0 = 2e^2/h$ = (12.9 kΩ). It is apparent that the sharp discontinuity discussed as an indication of the carboxylic oxygen flip-up motion is only observed above the carboxylic oxygen atoms. Jump of the conductance upon the approach to the clean Ag(111) surface was discussed in details in the ref. [6] and could be attributed to the formation of the contact between the tip and the surface. Zero of the z-scale was established according to the procedure of the absolute z value calibration described in the supplement file. **B,** Schematics of the contact formation process. Four stages of the contact formation correspond to the points marked on the approach curve in Fig. 2A.

**Figure 3 | Reactivity maps of PTCDA functional groups and artificial vacancy created in PTCDA layer. A,** STM image (5×5 nm$^2$, $I$ = 3 nA, $V_b$ = −15 mV) of the PTCDA/Ag(111) monolayer. Two non-equivalent molecules composing the unit-cell of the layer are marked as type A and B [14, 18, 40]. The image was taken at the distance of (2.4 ± 0.8) Å above the carboxylic oxygens (see the height calibration procedure in the supplement file). **B,** STM image (5×5 nm$^2$, I = 5.0 nA, Vb = −10 mV) taken at a distance of < 2.4 Å above the carboxylic oxygens. Contrary to the image in Fig. 3A here almost all carboxylic oxygen atoms respond. **C,** Vacancy island created by removing PTCDA molecules with the tip. Image 10×7 nm$^2$, $I$ = 0.1 nA, $V_b$ = −340 mV.

**Figure 4 | Statistics of approaches at the positions of the carboxylic oxygens of PTCDA. A,** Distribution of $I/V_b$-conductances in the tunnelling regime (curve 1) before the discontinuity, and immediately after the discontinuity (curve 2). Curve "max" – distribution of conductance maxima occurring for the strained contact, based on 146 retraction curves. Distributions were evaluated on dataset consisting of 282 approach curves (Fig. 5A), all carried out with identical tip. **B,** Distribution of tip-to-carboxylic-oxygen distances at which the discontinuity occurs; red – type A PTCDA, blue – type B PTCDA (In the commensurate superstructure of PTCDA on Ag(111) there are two inequivalent sites, A and B shown in Figs. 3A,B). Data (recorded with several different tips) are based on 110 approaches for type A and 115 approaches for type B. Green curve: distribution of tip-to-carboxylic-oxygen distances at which conductance maximum occurs (based on the 146 retraction curves of the dataset presented in Figs. 4A and 5A). For all z-calibrations in this Figure, cf. supplementary file.

**Figure 5 | Stretching statistics of the tip-PTCDA contact. A,** Collection of 282 retraction spectra recorded with identical tip. Brightness scales with frequency of events. For all z-calibrations in this Figure, cf. supplementary file. **B,** One of the approach retraction curves from the set shown in Fig. 5A. The curve visualizes typical conductance behaviour upon stretching of molecular contact.

**Figure 6 | Stretching various types of molecular contacts.** Approach retraction curves corresponding to the stretching of different molecular contacts. Black – stretching of the contact to PTCDA residing in the compact layer, results in PTCDA molecule to fall back after detachment from the tip. Red – stretching of the contact to PTCDA residing in the compact layer, results in PTCDA molecule to stay at the tip. Blue – stretching of the contact to PTCDA residing at the edge of the large island, results in the displacement of the molecule from its original position (also shown in the inset of Fig. 7B). Green – stretching of the contact to single PTCDA molecule, results in the displacement of the molecule from its original position (also shown in the inset of Fig. 7A). Zero of the z-scale was established according to the procedure of the absolute z value calibration described in the supplement file.

**Figure 7 | Conductance spectra of stretched transport junction. A, B,** 2 approach and retraction spectra of the junction recorded on a separate molecule (A, junction I) and a molecule at an island boundary (B, junction II) ($V_b$ = 2 mV). Zero of the z-scale was established according to the procedure of the absolute z value calibration described in the supplement file. Images show the investigated molecule before and after the experiment, the cross indicates the point of approach. Vertical bars: Range of conductance values from $G$-$V_b$-spectra in C and D. **C,** Series of $G$-$V_b$ -spectra (smoothed) measured at z-positions indicated in A by vertical bars ($V_{mod}$=4 mV, $\nu_{mod}$=723 Hz). **D,** Series of $G$-$V_b$-spectra (smoothed) measured at z-positions indicated in B by vertical bars ($V_{mod}$=4 mV, $\nu_{mod}$=723 Hz). Colors in C and D code the regimes in A and B.

**Figure 8 | Conductance simulation and structural model of sample contact cleavage. A, B** Color-coded plot of the transmission matrix T, calculated with NRG (cf. text), as a function of bias (vertical axis) and position of the transport orbital $\varepsilon_0$ relative to the Fermi level (horizontal axis), for lead couplings $\Gamma$=10 meV at 1K (A) and 20 meV at 9K (B). Electron correlation $U$=100meV in both cases. (A) illustrates the basic features to be expected when the transport orbital passes $E_F$, while (B) is close to our experimental parameters. Occupancy $n$ of the transport orbital is indicated in A. Within each of the two simulations, we have kept $\Gamma$ constant for clarity, although stretching the junction reduces $\Gamma$. However, it is easy to imagine how the plot would be modified by an increasing $\Gamma$ as we move away from the Kondo regime deeper into the n=2 regime. **C** Vertical cuts through the plots in A (red) and B (blue), at $\varepsilon_0$=-50meV. This shows the three-peak structure in the Kondo regime, with the singly occupied level at $\varepsilon_0$, the doubly occupied level at $\varepsilon_0$+U, and the pinned Kondo peak in the centre. Larger lead couplings (B) wash out the distinct peak structure of $\varepsilon_0$ and $\varepsilon_0$+$U$, but the physics is unchanged.

**Figure 9 | Structural model of sample contact cleavage.** Schematic structural model of the gradual $\pi$-bond cleavage, with qualitative evolution of the junction parameters, for further information cf. text.

References

1.      Nitzan, A. & Ratner, M.A. Electron transport in molecular wire junctions. Science 300, 1384-1389 (2003).

2.      Tao, N.J. Electron transport in molecular junctions. Nature Nanotechnology 1, 173-181 (2006).

3.      McCreery, R.L. Molecular electronic junctions. Chemistry of Materials 16, 4477-4496 (2004).

4.      Joachim, C., Gimzewski, J.K., Schlittler, R.R. & Chavy, C. Electronic Transparency of a Single C-60 Molecule. Phys. Rev. Lett. 74, 2102-2105 (1995).

5.      Moresco, F. Manipulation of large molecules by low-temperature STM: model systems for molecular electronics. Phys. Rep. 399, 175-225 (2004).

6.      Limot, L., Kroger, J., Berndt, R., Garcia-Lekue, A. & Hofer, W.A. Atom transfer and single-adatom contacts. Phys. Rev. Lett. 94, 126102 (2005).

7.      Kröger, J., Jensen, H. & Berndt, R. Conductance of tip-surface and tip-atom junctions on Au(111) explored by a scanning tunnelling microscope. New J. Phys. 9, 153-153 (2007).

8.      Neel, N. et al. Controlled contact to a C-60 molecule. Phys. Rev. Lett. 98, 065502 (2007).

9.      Neel, N., Kroger, J., Limot, L. & Berndt, R. Conductance of single atoms and molecules studied with a scanning tunnelling microscope. Nanotechnology 18, 3 (2007).

10.     Park, J. et al. Coulomb blockade and the Kondo effect in single-atom transistors. Nature 417, 722-725 (2002).

11.     Liang, W.J., Shores, M.P., Bockrath, M., Long, J.R. & Park, H. Kondo resonance in a single-molecule transistor. Nature 417, 725-729 (2002).

12.     Kubatkin, S. et al. Single-electron transistor of a single organic molecule with access to several redox states. Nature 425, 698-701 (2003).

13.     Bohringer, M. et al. Corrugation reversal in scanning tunneling microscope images of organic molecules. Phys. Rev. B 57, 4081-4087 (1998).

14.     Umbach, E., Glockler, K. & Sokolowski, M. Surface "architecture" with large organic molecules: interface order and epitaxy. Surf. Sci. 404, 20-31 (1998).

15.     Zou, Y. et al. Chemical bonding of PTCDA on Ag surfaces and the formation of interface states. Surf. Sci. 600, 1240-1251 (2006).

16.     Tautz, F.S. et al. Strong electron-phonon coupling at a metal/organic interface: PTCDA/Ag(111). Phys. Rev. B 65, 125405-125414 (2002).

17.     Eremtchenko, M., Schaefer, J.A. & Tautz, F.S. Understanding and tuning the epitaxy of large aromatic adsorbates by molecular design. Nature 425, 602-605 (2003).

18.     Kraft, A. et al. Lateral adsorption geometry and site-specific electronic structure of a large organic chemisorbate on a metal surface. Phys. Rev. B 74, 041402(R) (2006).

19.     Temirov, R., Soubatch, S., Luican, A. & Tautz, F.S. Free-electron-like dispersion in an organic monolayer film on a metal substrate. Nature 444, 350-353 (2006).

20.     Bannani, A., Bobisch, C. & Moller, R. Ballistic Electron Microscopy of Individual Molecules. Science 315, 1824-1828 (2007).

21.     Glockler, K. et al. Highly ordered structures and submolecular scanning tunnelling microscopy contrast of PTCDA and DM-PBDCI monolayers on Ag(111) and Ag(110). Surf. Sci. 405, 1-20 (1998).

22.     Hauschild, A. et al. Molecular distortions and chemical bonding of a large pi-conjugated molecule on a metal surface. Phys. Rev. Lett 94, 036106-036109 (2005).

23.     Hauschild, A. et al. Comment on "Molecular distortions and chemical bonding of a large pi-conjugated molecule on a metal surface" - Reply. Phys. Rev. Lett 95, 209602 (2005).

24.     Rurali, R., Lorente, N. & Ordejon, P. Comment on "Molecular distortions and chemical bonding of a large pi-conjugated molecule on a metal surface". Phys. Rev. Lett 95, 209601 (2005).

25.     Kondo, J. Resistance Minimum in Dilute Magnetic Alloys. Progress of Theoretical Physics 32, 37-49 (1964).

26.     Grobis, M., Rau, I.G., Potok, R.M. & Goldhaber-Gordon, D. Kondo Effect in Mesoscopic Quantum Dots. arXiv:cond-mat/0611480v1 (2006).

27.     Wingreen, N.S. & Meir, Y. Anderson Model out of Equilibrium - Noncrossing-Approximation Approach to Transport through a Quantum-Dot. Phys.Rev.B 49, 11040-11052 (1994).

28.     Goldhaber-Gordon, D. et al. From the Kondo regime to the mixed-valence regime in a single-electron transistor. Phys. Rev. Lett. 81, 5225-5228 (1998).

29.     Goldhaber-Gordon, D. et al. Kondo effect in a single-electron transistor. Nature 391, 156-159 (1998).

30.     Cronenwett, S.M., Oosterkamp, T.H. & Kouwenhoven, L.P. A tunable Kondo effect in quantum dots. Science 281, 540-544 (1998).

31.    Nygard, J., Cobden, D.H. & Lindelof, P.E. Kondo physics in carbon nanotubes. Nature 408, 342-346 (2000).

32.    Jarillo-Herrero, P. et al. Orbital Kondo effect in carbon nanotubes. Nature 434, 484-488 (2005).

33.    Li, J.T., Schneider, W.D., Berndt, R. & Delley, B. Kondo scattering observed at a single magnetic impurity. Physical Review Letters 80, 2893-2896 (1998).

34.    Madhavan, V., Chen, W., Jamneala, T., Crommie, M.F. & Wingreen, N.S. Tunneling into a single magnetic atom: Spectroscopic evidence of the Kondo resonance. Science 280, 567-569 (1998).

35.    Wahl, P. et al. Kondo temperature of magnetic impurities at surfaces. Phys. Rev. Lett. 93, 176603 (2004).

36.    Nagaoka, K., Jamneala, T., Grobis, M. & Crommie, M.F. Temperature dependence of a single Kondo impurity. Phys. Rev. Lett. 88, - (2002).

37.    Neel, N. et al. Conductance and Kondo effect in a controlled single-atom contact. Phys. Rev. Lett. 98, 016801 (2007).

38.    Zhao, A. et al. Controlling the Kondo Effect of an Adsorbed Magnetic Ion Through Its Chemical Bonding. Science 309, 1542-1544 (2005).

39.    Hofer, W.A., Fisher, A.J., Wolkow, R.A. & Grutter, P. Surface relaxations, current enhancements, and absolute distances in high resolution scanning tunneling microscopy. Phys. Rev. Lett. 87, 236104 (2001).

40.    Rohlfing, M., Temirov, R. & Tautz, F.S. Structural properties and scanning tunneling data of a monolayer of 3,4,9,10-perylene-tetracarboxylic-dianhydride (PTCDA) on the Ag(111) surface. (submitted).

41.    Kilian, L. et al. Structural and electronic phase transition at a metal-organic interface: The role of intermolecular interactions. (submitted).

42.    Kergueris, C. et al. Electron transport through a metal-molecule-metal junction. Phys.Rev.B 59, 12505-12513 (1999).

43.    Reed, M.A., Zhou, C., Muller, C.J., Burgin, T.P. & Tour, J.M. Conductance of a molecular junction. Science 278, 252-254 (1997).

44.    Parks, J.J. et al. Tuning the Kondo effect with a mechanically controllable break junction. arXiv:cond-mat/0610555 v1 (2006).

45.    Norskov, J.K. Chemisorption on Metal-Surfaces. Rep. Prog. Phys. 53, 1253-1295 (1990).

46.    Anderson, P.W. Localized Magnetic States in Metals. Phys. Rev. 124, 41 (1961).

47.    Newns, D.M. Self-Consistent Model of Hydrogen Chemisorption. Phys. Rev. 178, 1123 (1969).

48.    Costi, T.A., Hewson, A.C. & Zlatic, V. Transport-Coefficients of the Anderson Model Via the Numerical Renormalization-Group. J. Phys. Cond.Mat. 6, 2519-2558 (1994).

49.    Schiller, A. & Hershfield, S. Theory of scanning tunneling spectroscopy of a magnetic adatom on a metallic surface. Phys.Rev.B 61, 9036 (2000).

50.    Krishna-murthy, H.R., Wilkins, J.W. & Wilson, K.G. Renormalization-group approach to the Anderson model of dilute magnetic alloys. I. Static properties for the symmetric case. Phys.Rev.B 21, 1003 (1980).

51.    Meir, Y. & Wingreen, N.S. Landauer formula for the current through an interacting electron region. Phys. Rev. Lett. 68, 2512 (1992).

52.    For U/G<1 or a strongly asymmetric junction, the transmission matrix T(w) remains independent of the bias, while in a symmetric junction non-equilibrium effects to the T-matrix have to be taken into account at large bias.

53.    Peters, R., Pruschke, T. & Anders, F.B. Numerical renormalization group approach to Green's functions for quantum impurity models. Phys.Rev.B 74, 245114-245112 (2006).

54.    Weichselbaum, A. & Delft, J.v. Sum-rule Conserving Spectral Functions from the Numerical Renormalization Group. arXiv:cond-mat/0607497v1 (2006).

55.    Langreth, D.C. Friedel Sum Rule for Anderson's Model of Localized Impurity States. Phys.Rev. 150, 516 (1966).

56.    Yu, L.H. & Natelson, D. The Kondo Effect in C60 Single-Molecule Transistors. Nano Lett. 4, 79-83 (2004).

**A**

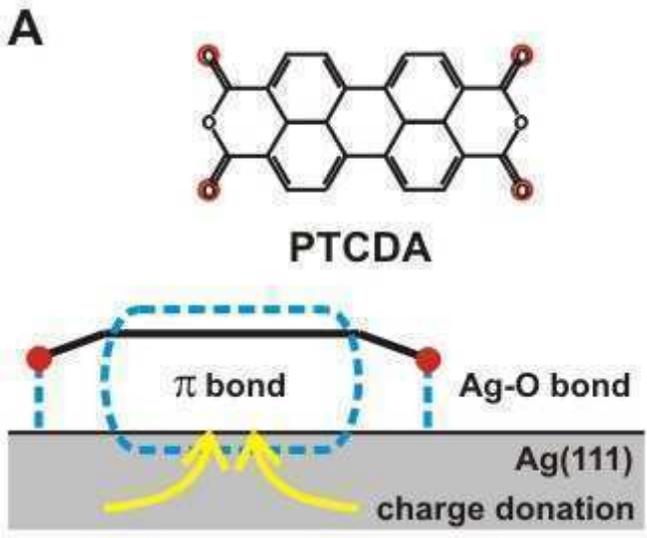

PTCDA

**B**

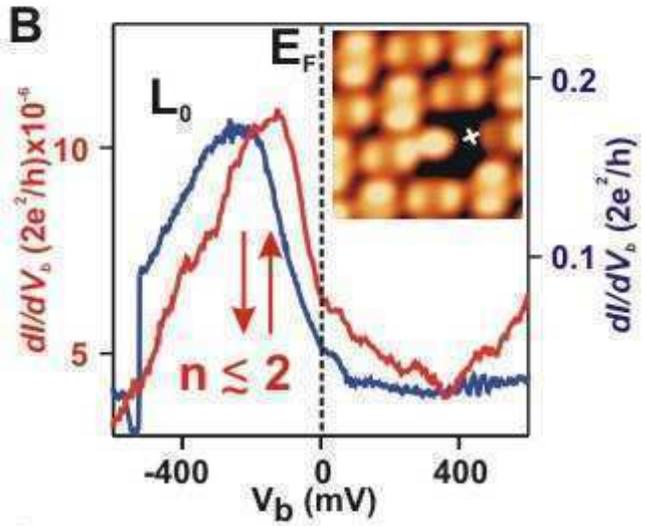

**C**

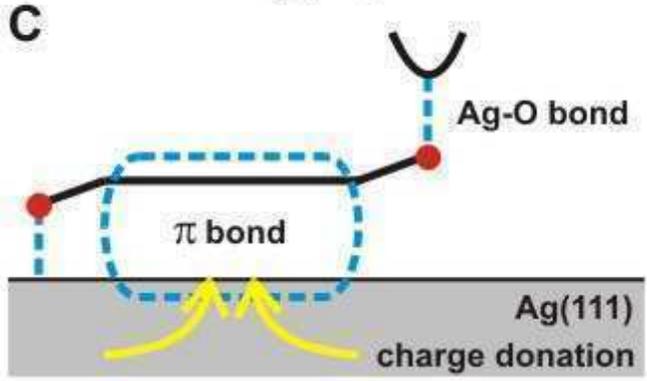

Temirov, Lassise,
Anders, Tautz (2007),
Figure 1

**A**

**B**

Temirov, Lassise, Anders & Tautz (2007), Figure 2

**A**

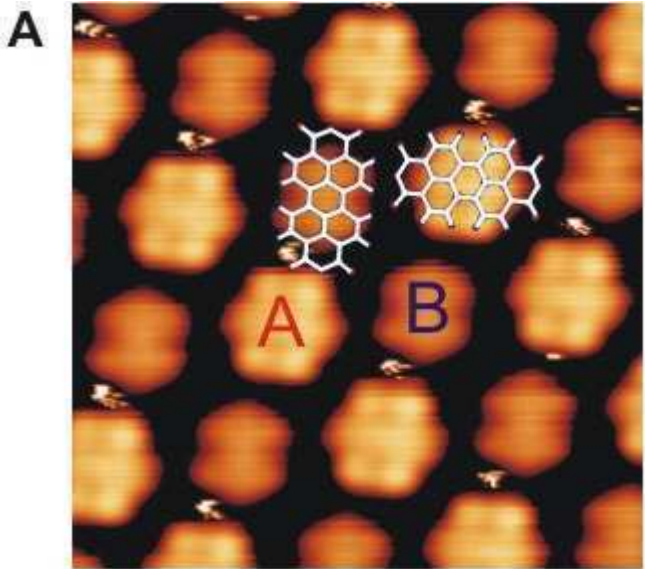

**B**

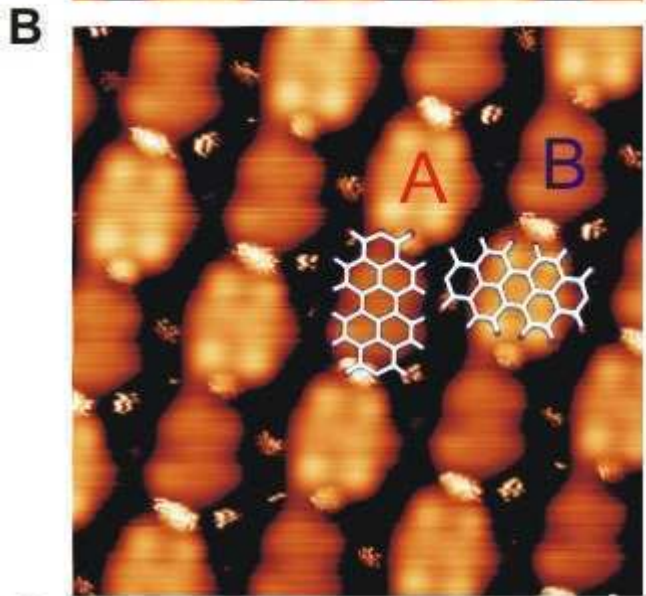

**C**

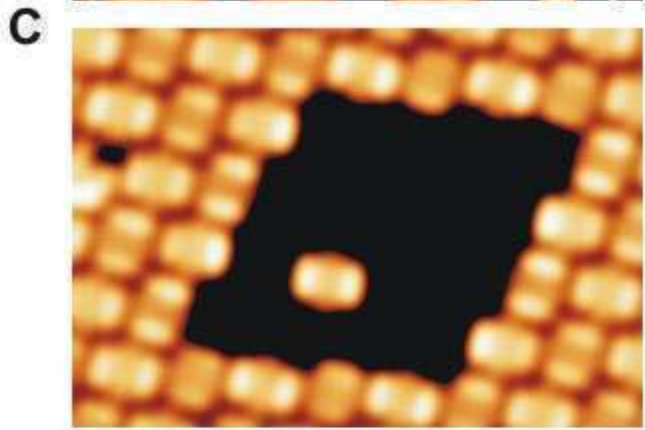

Temirov, Lassise,
Anders, Tautz (2007),
Figure 3

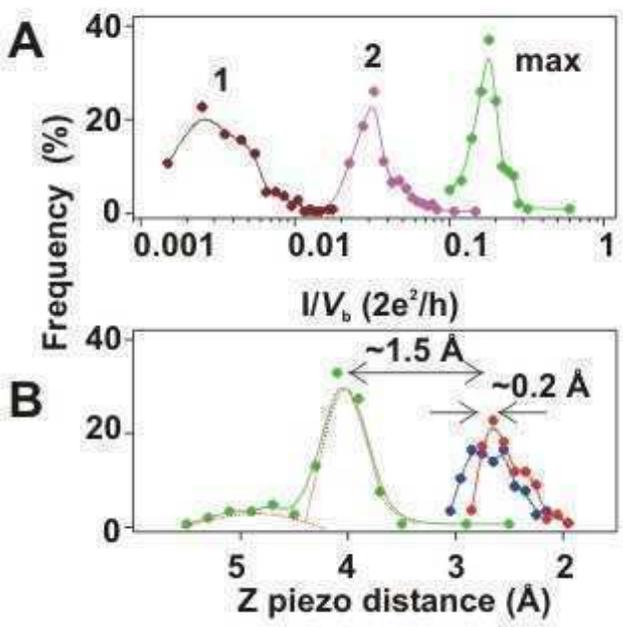

**Temirov, Lassise, Anders, Tautz (2007), Figure 4**

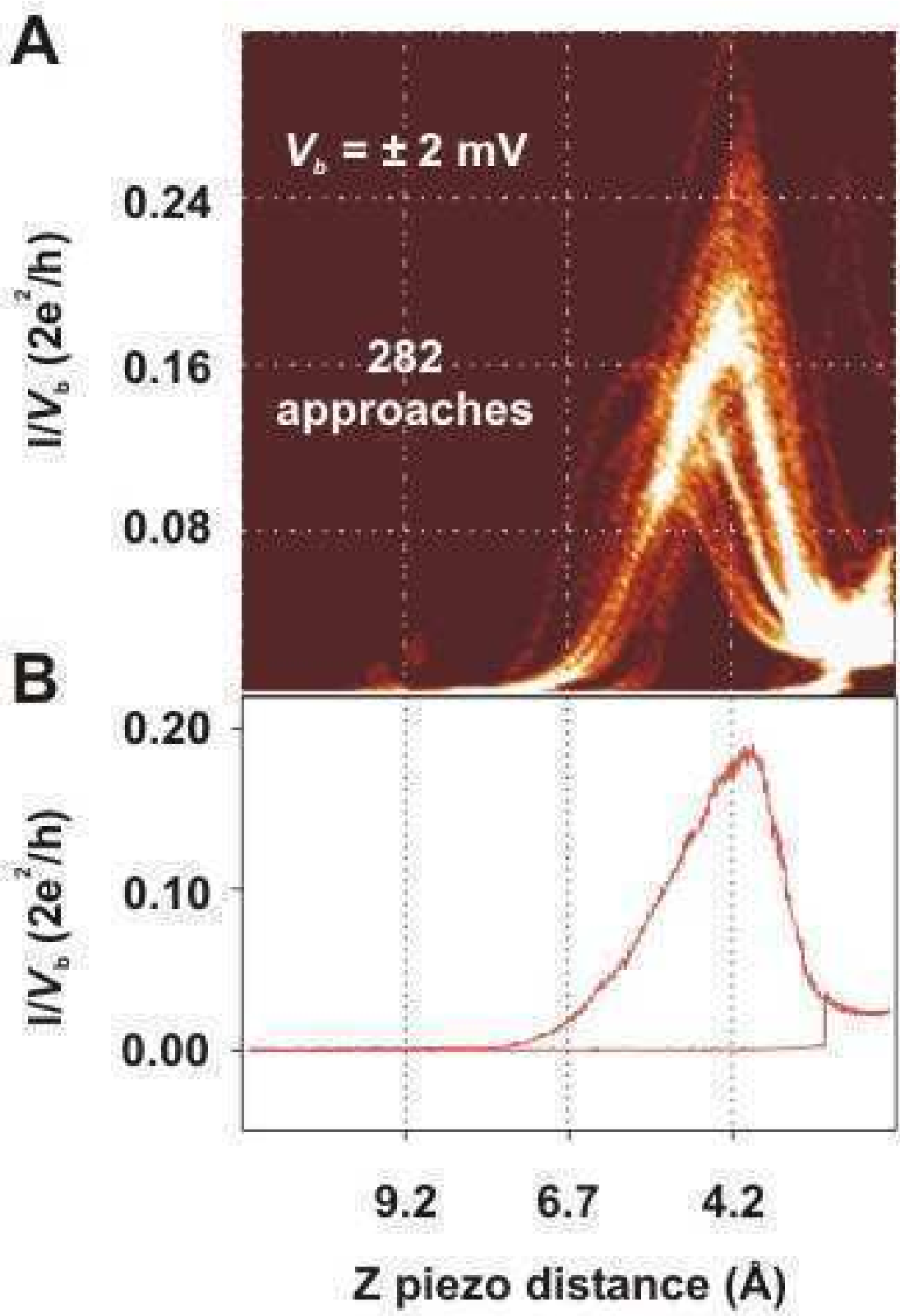

Temirov, Lassise, Anders, Tautz (2007), Figure 5

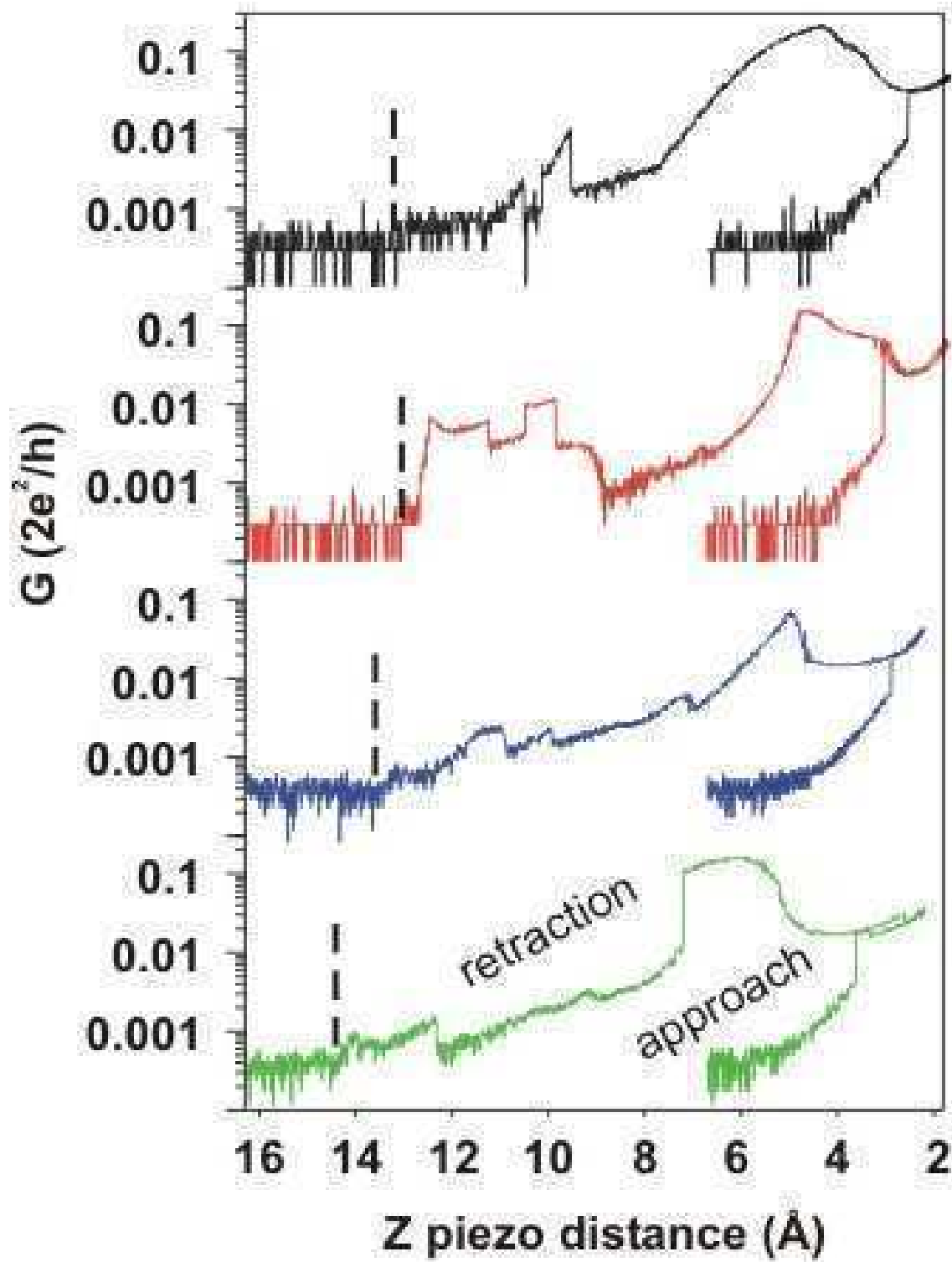

G (2e²/h)

retraction

approach

Z piezo distance (Å)

**Temirov, Lassise, Anders, Tautz (2007), Figure 6**

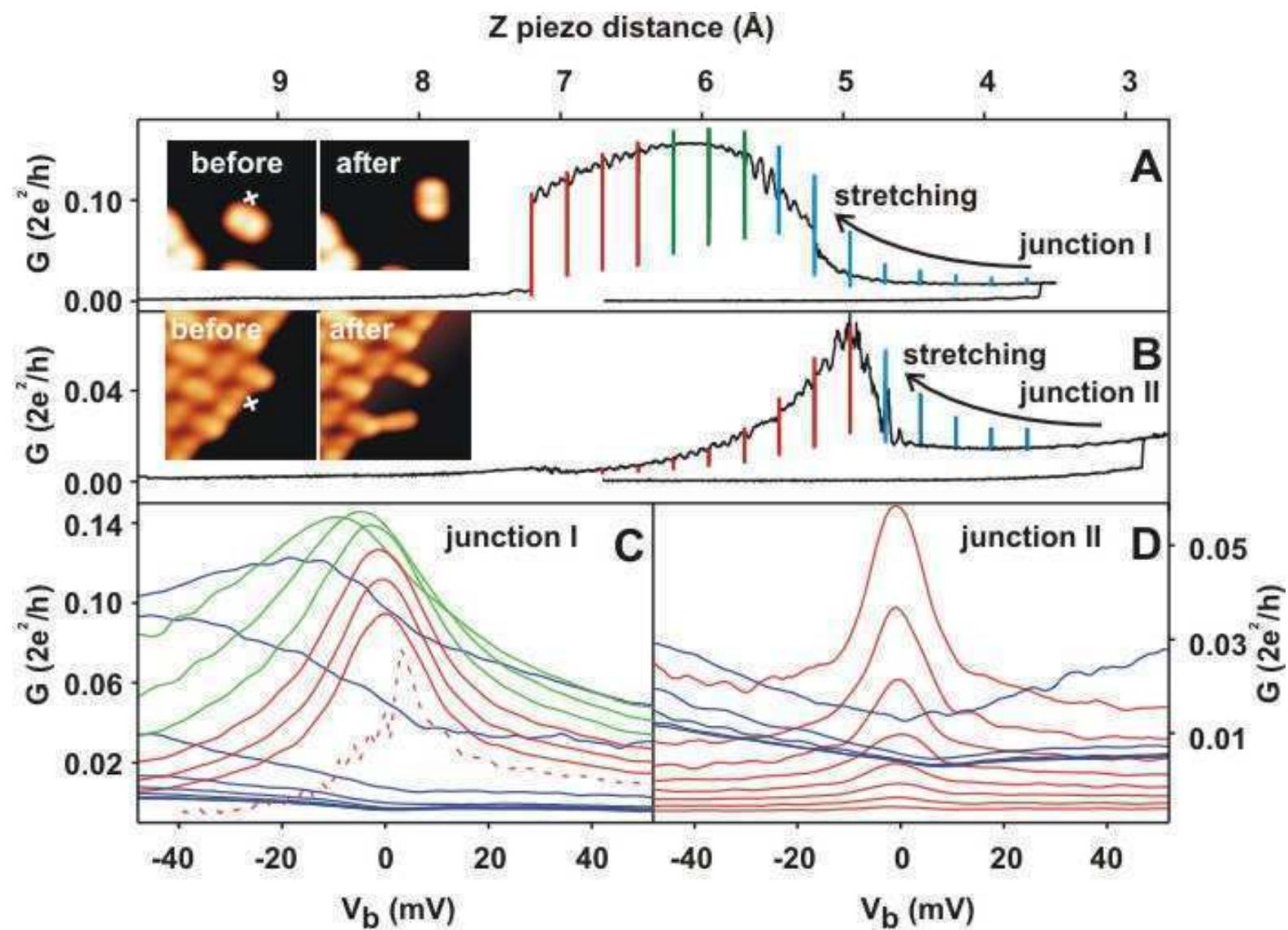

**Temirov, Lassise, Anders & Tautz (2007), Figure 7**

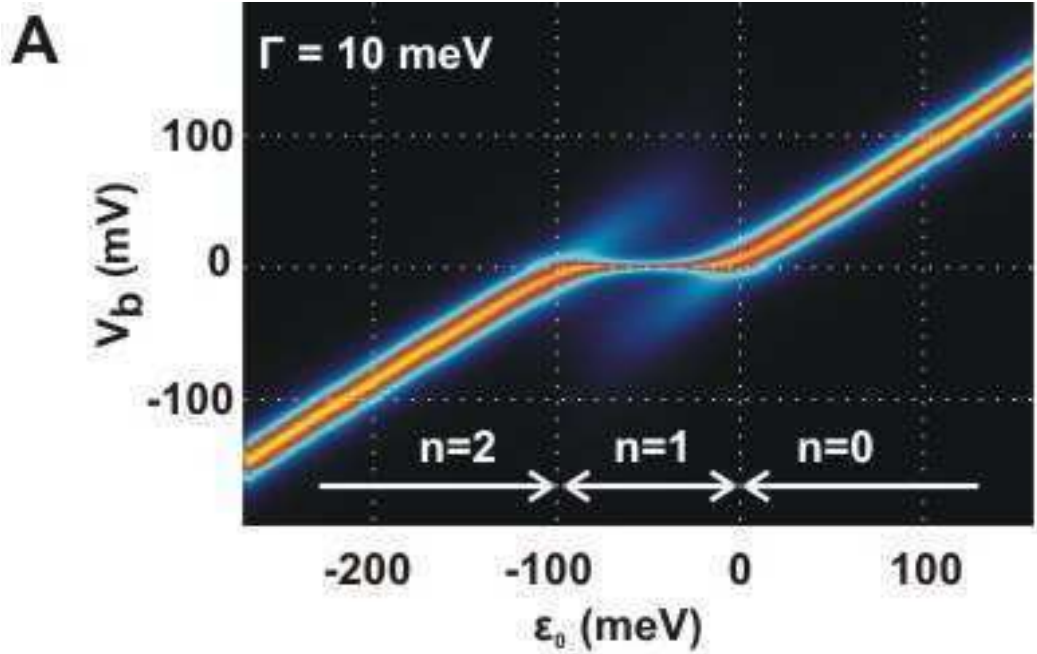

**A** Γ = 10 meV

V_b (mV)

n=2    n=1    n=0

ε_0 (meV)

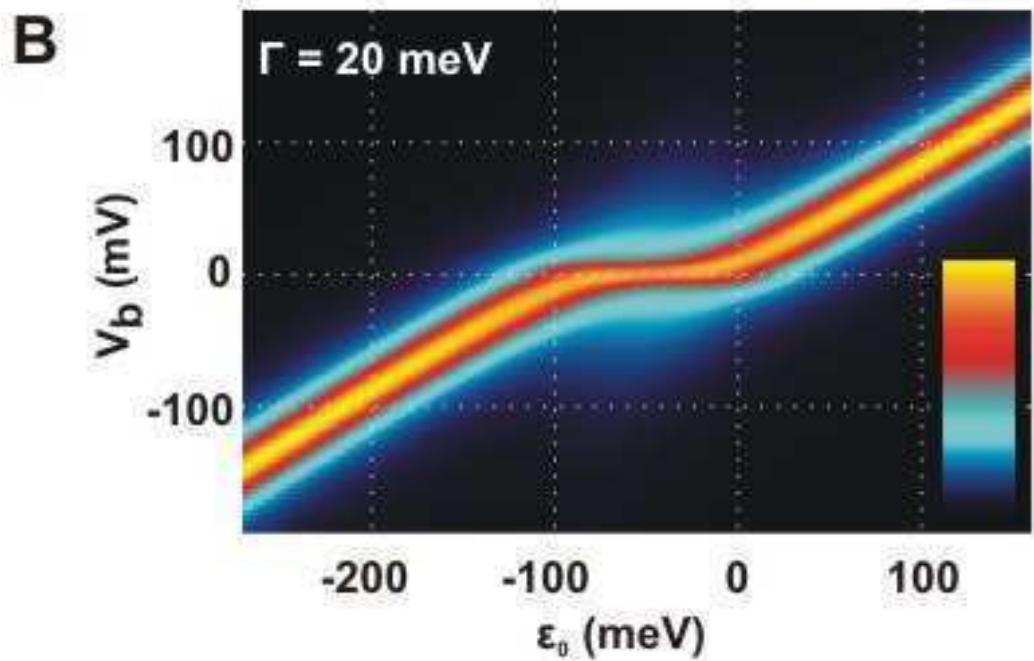

**B** Γ = 20 meV

V_b (mV)

ε_0 (meV)

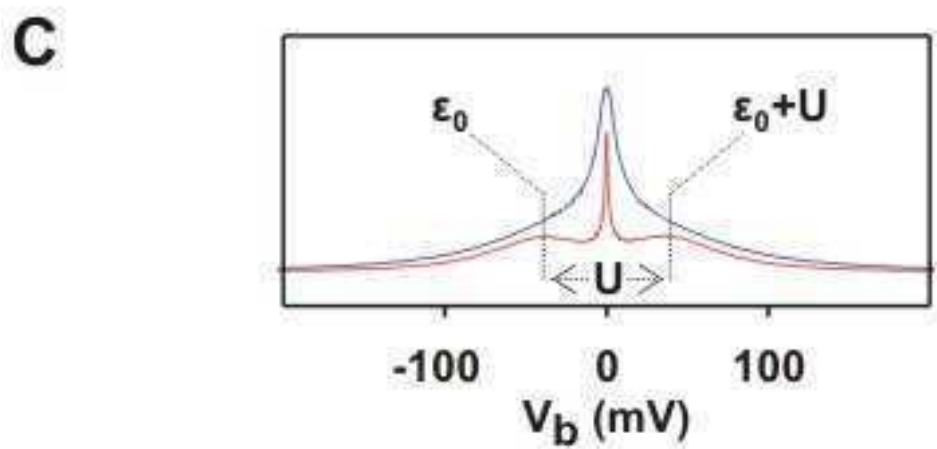

**C**

ε_0    ε_0+U

U

V_b (mV)

Temirov, Lassise, Anders,
Tautz (2007), Figure 8

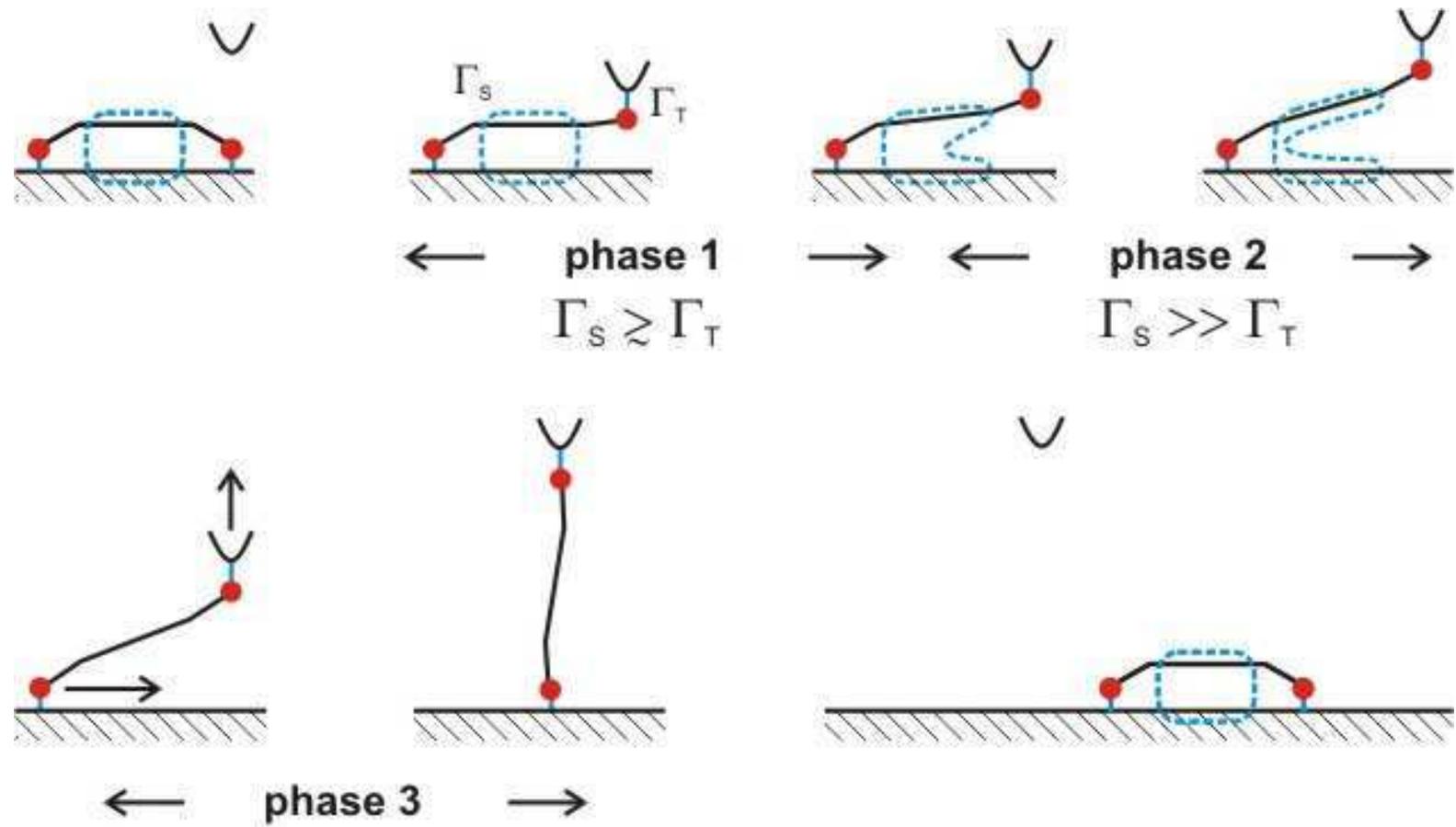

Temirov, Lassise, Anders & Tautz (2007), Figure 9

# Supplementary Information

## 1. Absolute z-calibration of approach spectra

*Details of the calibration procedure*

Once the tip has been placed above the Ag surface at $I$ = 0.1 nA, $V_b$ = −340 mV (completion of step 2, cf. Methods section), it is moved towards the Ag surface until contact is made (step 3, cf. Methods section). A typical approach curve above Ag is shown in Supplementary Fig. 1a. The step at (6.3 ± 0.2) Å indicates that contact has been made. In the present example, the conductance after the step is 1.7 $G_0$. Values in this range, and in particular larger than $G_0$, are common for tip contacts with flat surfaces (S1). This proves that more than one conduction channel has been established, indicating in turn that the contact is not formed between just the single apex atom and a single atom in the surface. The contact appears after a piezo-shift of (6.3 ± 0.2) Å (Supplementary Fig. 1a). This value, however, must be corrected, because the attractive forces between tip and surface distort both of them before the contact is made. This distortion has been calculated quantitatively in ref. S1, with the result that the tip and sample move towards each other by 1 to 2 Å; contact is hence made at a value on the piezo-scale for which the undistorted tip would still be 1 to 2 Å away from the undistorted surface. When seeking to determine the distance of a tip to the lattice planes of a surface, in the regime when the tip is far away from the surface and therefore both are undistorted, we therefore have to add 1 to 2 Å to the piezo-shift determined from an approach curve such as presented in Supplementary Fig. 1a. The tip-to-sample distance in Supplementary Fig. 1b is therefore (6.3 + 1.5 ± 0.5) Å = (7.8 ± 0.5) Å. Having accounted for the distortion effect by the (+1.5 ± 0.5) Å correction, we can from now on disregard the distortion of tip and sample in our discussion.

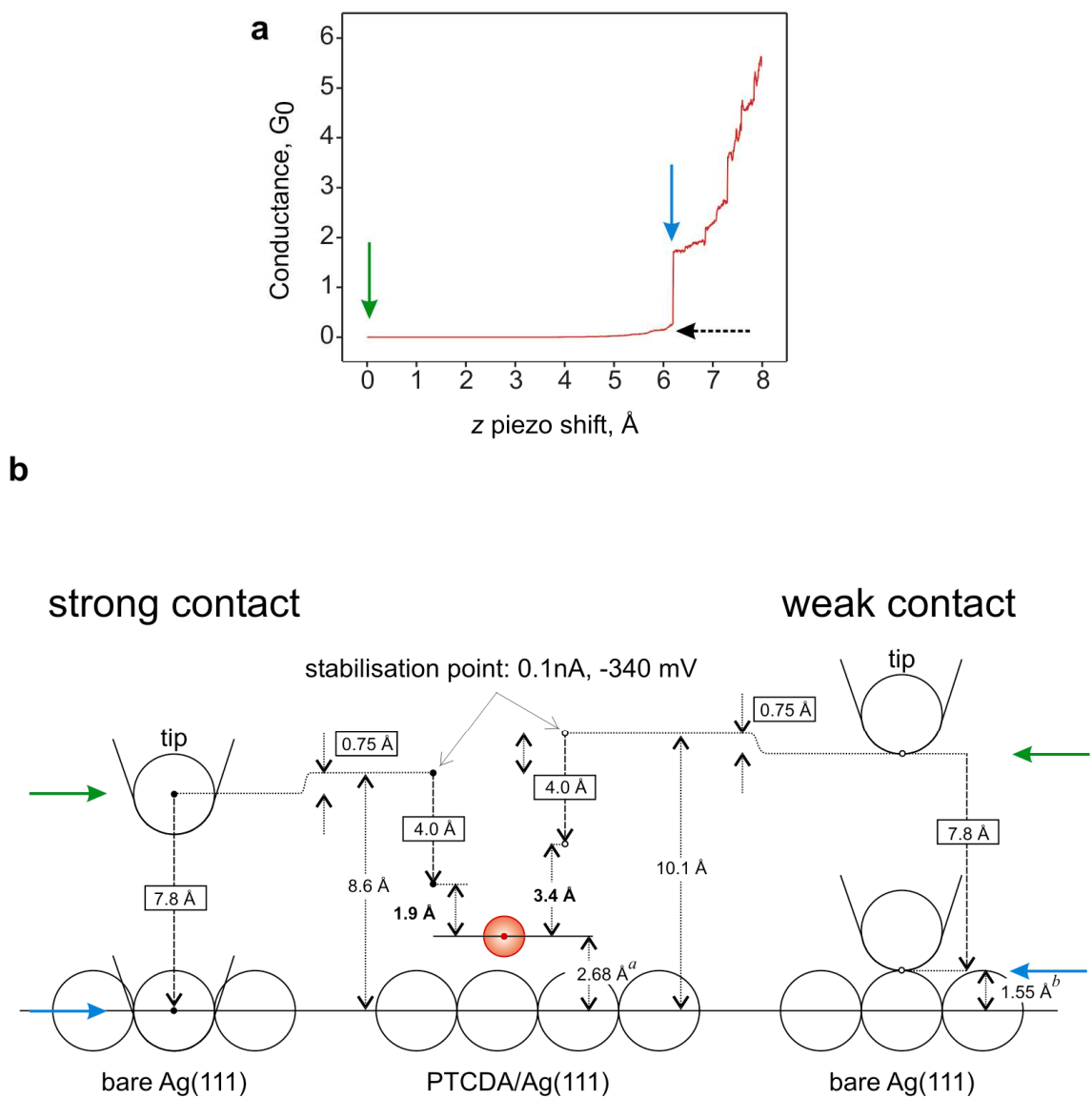

**Supplementary Figure 1 | Approach spectrum above Ag(111) and *z*-calibration. a**, Approach spectrum measured with a bias of $V_b = -2$ mV. Before opening the feedback loop, the tip was stabilized at $I = 0.1$ nA and $V_b = -340$ mV. Dashed arrow indicates the total distortion of the tip-sample contact. Blue and green arrows correspond the position marked in b. **b**, Scale model of the *z*-calibration procedure. Left and right panels: Approach above Ag(111) (cf. Methods, step 3), assuming weak and strong contacts. Middle panel: Approach above carboxylic oxygen atoms, the latter represented by a red circle with covalent radius of $r_c = 0.7$ Å (ref. S2). Tip and substrate atoms are represented by open circles with the covalent radius of silver ($r_c = 1.55$ Å) (ref. S2). Framed distances are piezo-shifts. [a] The experimental value has been taken from ref. 18. [b] The value has been taken from ref. S3. Further discussion in the text of the Supplementary Information, section b.

The next issue is the atomic structure of the contact at the point of the conductance jump in Supplementary Fig. 1a. In Supplementary Fig. 1b we consider two limits: On the right, a weak contact is shown, assuming that the current jump occurs when the apex atom goes into first mechanical contact with the surface. On the left, the strong contact supposes that the apex atom becomes completely embedded in the first atomic layer of the surface. These contacts are extreme limits. Reality will lie somewhere in between. The conductance value of 1.7 $G_0$ indicates that the contact must be stronger than the weak contact on right of Supplementary Fig. 1b. In the following, we will base our calculations on both scenarios and interpolate between them. According to the Figure, they lead to an absolute tip height above Ag at the stabilisation point of (9.4 ± 1.6) Å. This corresponds to (6.7 ± 1.6) Å above the carboxylic oxygen atoms. The uncertainty of 1.6 Å originates from considering two extreme (and unlikely) limits of tip-surface contacts.

In the middle panel of Supplementary Fig. 1b the approach towards the carboxylic oxygen (displayed in red) is shown schematically. The piezo-shift to the point on the piezo-scale where the flip occurs is 4.0 Å, yielding 1.9 or 5.0 Å [=(3.4+1.6) Å] as the centre-to centre distance of the tip apex atom to the carboxylic oxygen just before the flip. From the distortion of the molecule (the carboxylic oxygen is 0.2Å below the average carbon position) we estimate an oxygen movement by ~ 0.5 Å to a symmetric position above the plane of the molecule. This yields an estimated Ag-O bonding distance of (2.9 ± 1.6) Å. The uncertainty in this number is given by the two limits of a strong and weak tip-Ag contact during $z$-calibration; the most likely value, corresponding to an 'intermediate' tip-Ag contact is indeed the central value of 2.9 Å. In reality, the (statistical) error bar will be much smaller; we estimate an error of ± 0.8 Å for the range of realistic tip-metal contacts. The Ag-O bond distance of (2.9 ± 0.8) Å is in reasonable agreement with known distances of Ag-O bonds, see above.

## 2. Supplementary Data

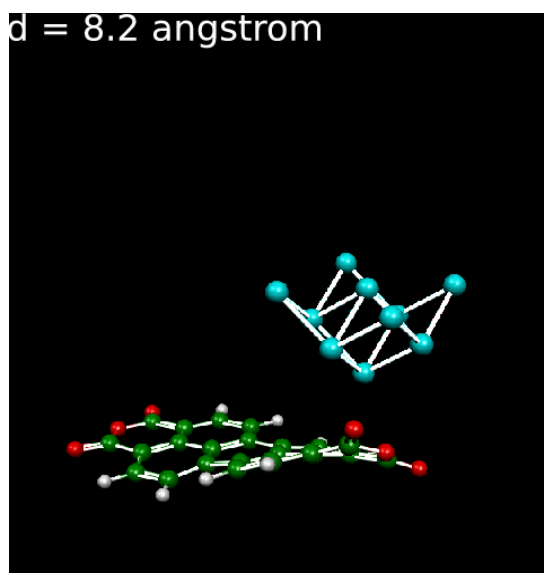

**Supplementary Figure 4 |** DFT calculated PTCDA-tip contact, with the tip 8.2 Å above the Ag(111) surface (not shown), i.e. after tip-retraction of 3.2 Å. Figure courtesy to M. Rohlfing, G. Cuniberti and B. Song.

**References**

S1. Limot, L., Kröger, J., Berndt, R., Garcia-Lekue, A., & Hofer, W.A. Atom transfer and single-adatom contacts. *Phys. Rev. Lett.* **94**, 126102(4) (2005).

S2.  www.webelements.com

S3. Henze, S. K. M., Bauer, O., Lee, T.-L., Sokolowski, M., Tautz, F. S. Vertical bonding distances of PTCDA on Au(1 1 1) and Ag(1 1 1): Relation to the bonding type. Surf. Sci. **601**, 1566-1573 (2007).